\documentclass[conference]{IEEEtran}
\usepackage[letterpaper]{geometry}
\newgeometry{left=1.57cm,right=1.57cm,bottom=2.54cm,top=2.0cm} %ORIGINAL
\usepackage{setspace}
\IEEEoverridecommandlockouts
\usepackage{cite}
\usepackage{amsmath,amssymb,amsfonts}
\usepackage{algorithmic}
\usepackage{graphicx}
\usepackage{textcomp}
\usepackage{float}
\usepackage{placeins}
\usepackage{multirow}
\usepackage{pifont}% http://ctan.org/pkg/pifont
\usepackage[table,xcdraw]{xcolor}
\begin{document}
	
\sloppy	
	\title{ApproxCS: Near-Sensor Approximate Compressed Sensing for IoT-Healthcare Systems}
	
	\author{\IEEEauthorblockN{Ayesha Siddique\IEEEauthorrefmark{1},
			Osman Hasan\IEEEauthorrefmark{1},
			Faiq Khalid\IEEEauthorrefmark{2}, 
			Muhammad Shafique\IEEEauthorrefmark{2}}
		\IEEEauthorblockA{\IEEEauthorrefmark{1}School of Electrical Engineering and Computer Sciences, \\ National University of Sciences and Technology, Islamabad, Pakistan}
		\IEEEauthorblockA{\IEEEauthorrefmark{2}Department of Computer Engineering, Vienna University of Technology, Vienna, Austria}
		\IEEEauthorblockA{Email: \{14mseeasiddique,osman.hasan\}@seecs.edu.pk, \{muhammad.shafique, faiq.khalid\}@tuwien.ac.at}
	}

	\maketitle
%========================================================================================================	
	\begin{abstract}
	Internet of Things (IoTs) is an emerging trend that has enabled an upgrade in the design of wearable healthcare monitoring systems through the (integrated) edge, fog, and cloud computing paradigm. Energy efficiency is one of the most important design metrics in such IoT-healthcare systems especially, for the edge and fog nodes. Due to the sensing noise and inherent redundancy in the input data, even the most safety-critical biomedical applications can sometimes afford a slight degradation in the output quality. Hence, such inherent error tolerance in the  bio-signals can be exploited to achieve high energy savings through the emerging trends like, the Approximate Computing which is applicable at both software and hardware levels. In this paper, we propose to leverage the approximate computing in digital Compressed Sensing (CS), through low-power approximate adders (LPAA) in an accurate Bernoulli sensing based CS acquisition (BCS). We demonstrate that approximations can indeed be safely employed in IoT-healthcare without affecting the detection of critical events in the biomedical signals. Towards this, we explored the trade-off between energy efficiency and output quality using the state-of-the-art ${\mathrm{l}_\mathrm{p}}^\mathrm{2d}$ RLS reconstruction algorithm. The proposed framework is validated with the MIT-BIH Arrhythmia database. Our results demonstrated approximately 59\% energy savings as compared to the accurate design.
	\end{abstract}
%========================================================================================================	

%========================================================================================================

\section{Introduction}
\begin{figure}[!t]
		\centering
		\vspace{-0.01in}
		\includegraphics[width=1\linewidth]{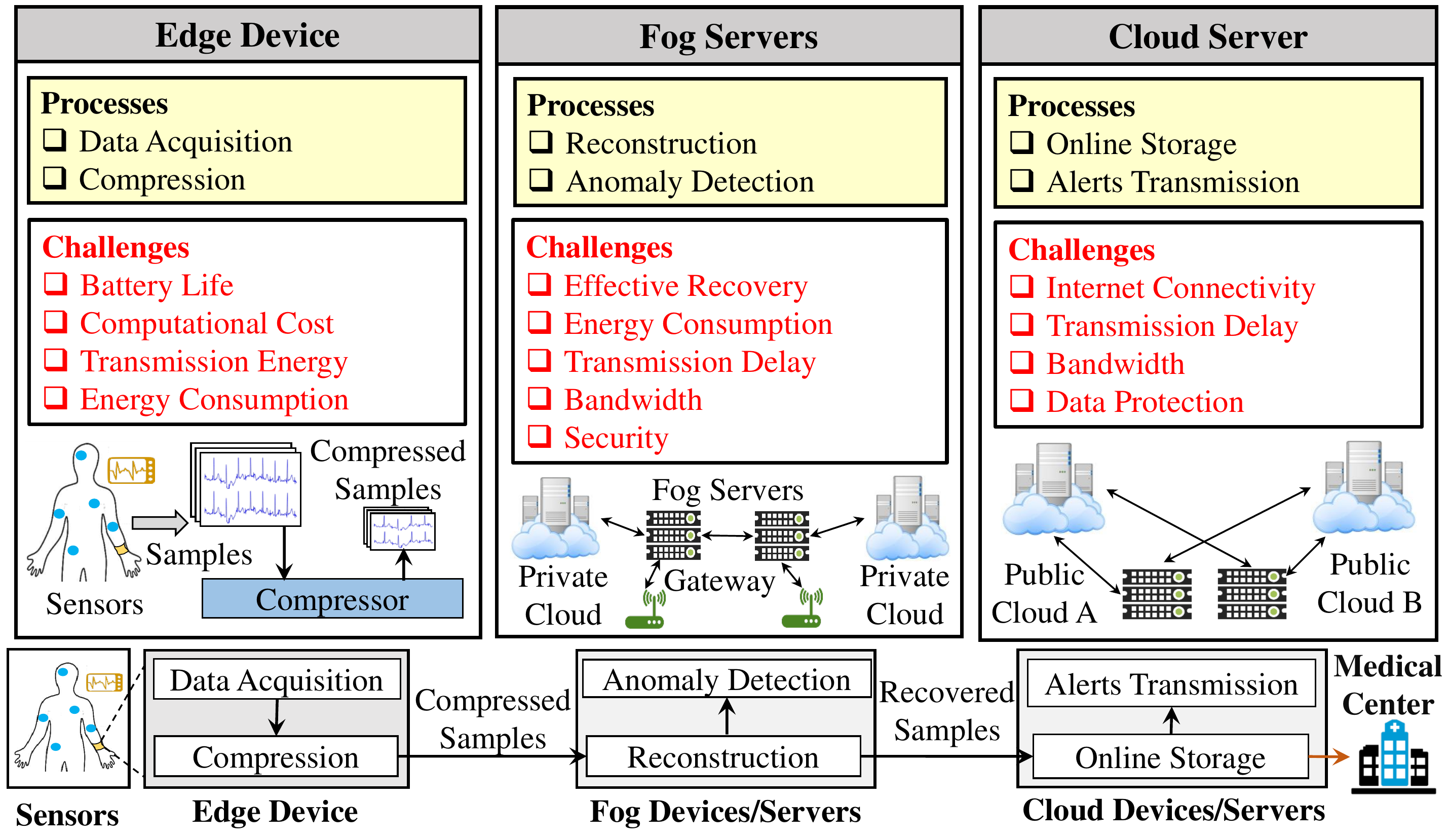}
		\caption{Tasks and Challenges in IoT based WBAN}
		\label{fig:allcomputations}
		\vspace{-0.1in}
	\end{figure}

	Internet of Things (IoT) is a megatrend in future technologies of the healthcare spectrum \cite{hassanalieragh2015health} where their economic worth is expected to be projected at \$1.9 trillion by 2020 \cite{gartner}. Such anticipated advancement has paved the way for novel system architectures in wireless body area network (WBAN), which aim to improve the currently deployed cloud computing paradigm \cite{varghese2018next}. Usually, the body sensors upload unprecedented amount of diverse physiological data (e.g., EEG, ECG, EMG, etc.) to centralized cloud servers for big data analytics. However, the cloud infrastructure poses a central bottleneck for such huge volume of data due to the high bandwidth requirement and unreliable Internet connection (especially, in under-developed countries). It is expedient to realize the distributed computing with computational offloading to more powerful layers, such as heterogeneous fogs \cite{zhang2015cloud}. However, it requires data transmission over a gateway, where low power wireless connection and slightly large bandwidth allocation are the major issues \cite{samie2018distributed}. This demand can be reduced by using state-of-the-art compression algorithms \cite{hilton1997wavelet}\cite{lu2000wavelet}\cite{mamaghanian2011compressed}\cite{craven2016energy} at edge devices, but they are sometimes computationally intensive. The near-sensor devices are either battery operated or they rely on energy-harvesting. These limited energy resources tend to reduce their processing capability. Recently, many ultra-low power processors \cite{wang2015energy}\cite{ma2015architecture}, hybrid \cite{bortolotti2014approximate} or non-volatile memories \cite{ghoneim2015review} and system-on-chips (SoCs) with integrated radio transceivers \cite{dazzi2018sub} have been developed. However, they contribute to quite meager energy efficiency gains on edge devices for big data analytics. Thus, providing low bandwidth within a few milliwatts computing power envelope is an open research challenge. Fig. \ref{fig:allcomputations} provides a sketch of the distributed processing in IoT based WBAN along with its associated challenges at each layer. 
	
%%%%%%%%%%%%%%%%%%%%%%%%%%%%%%%%%%%%%%%%%%%%%%%%%%%%%%%%%%%%%%%%%%%%%%%%%%%%%%%%%

\subsection{State-of-the-Art and Open Research Problems}
\begin{figure}[!t]
	\centering
	\vspace{-0.1in}
	\includegraphics[width=1\linewidth]{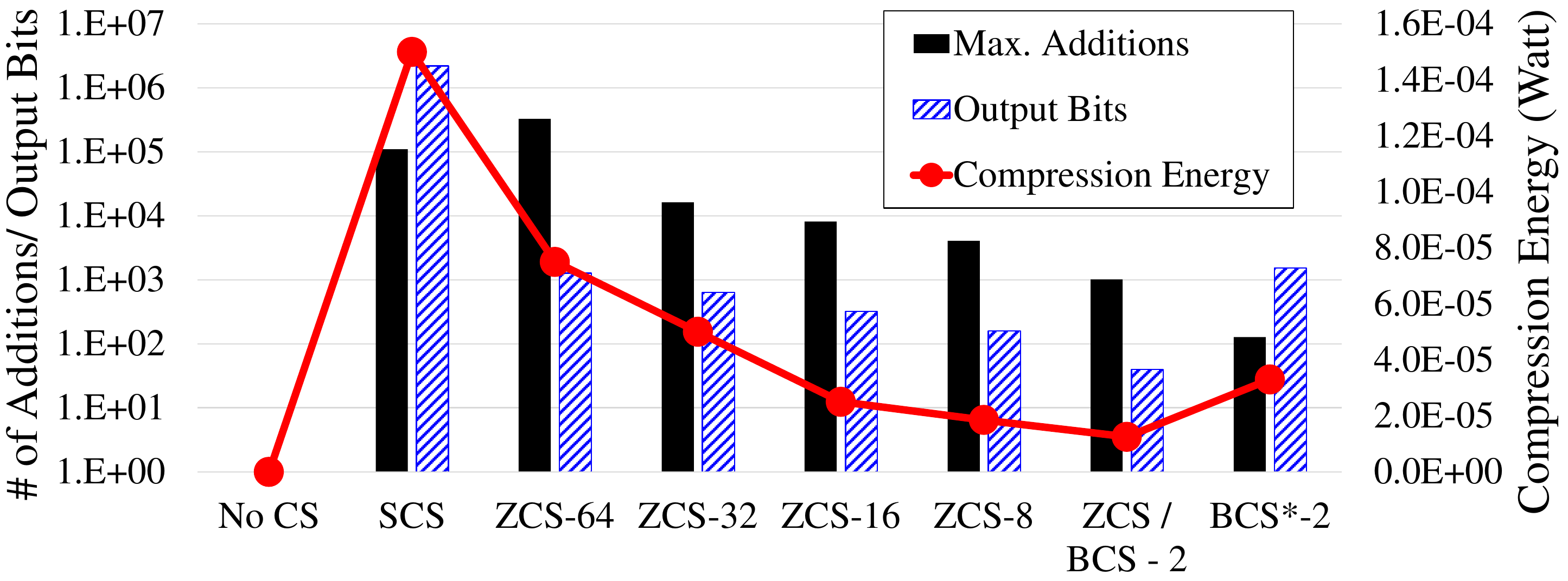}
	\caption{Comparison of the ${\mathrm{BCS}}^\mathrm{*}$ with other state-of-the-art CS acquisition techniques, using sensing matrix MxN with \textit{R} number of ones, i.e. SCS (215$\times$512), ZCS-64 (175$\times$512), ZCS-32 (192$\times$512), ZCS-16 (200$\times$512), ZCS-8 (209$\times$512), ZCS-2 (218$\times$512) and BCS-2 (128$\times$256)}
	\label{fig:comparitiveanalysis2}
	\vspace{-0.2in}
\end{figure}

\begin{figure}[!h]
	\centering
	\vspace{-0.1in}
	\includegraphics[width=1\linewidth]{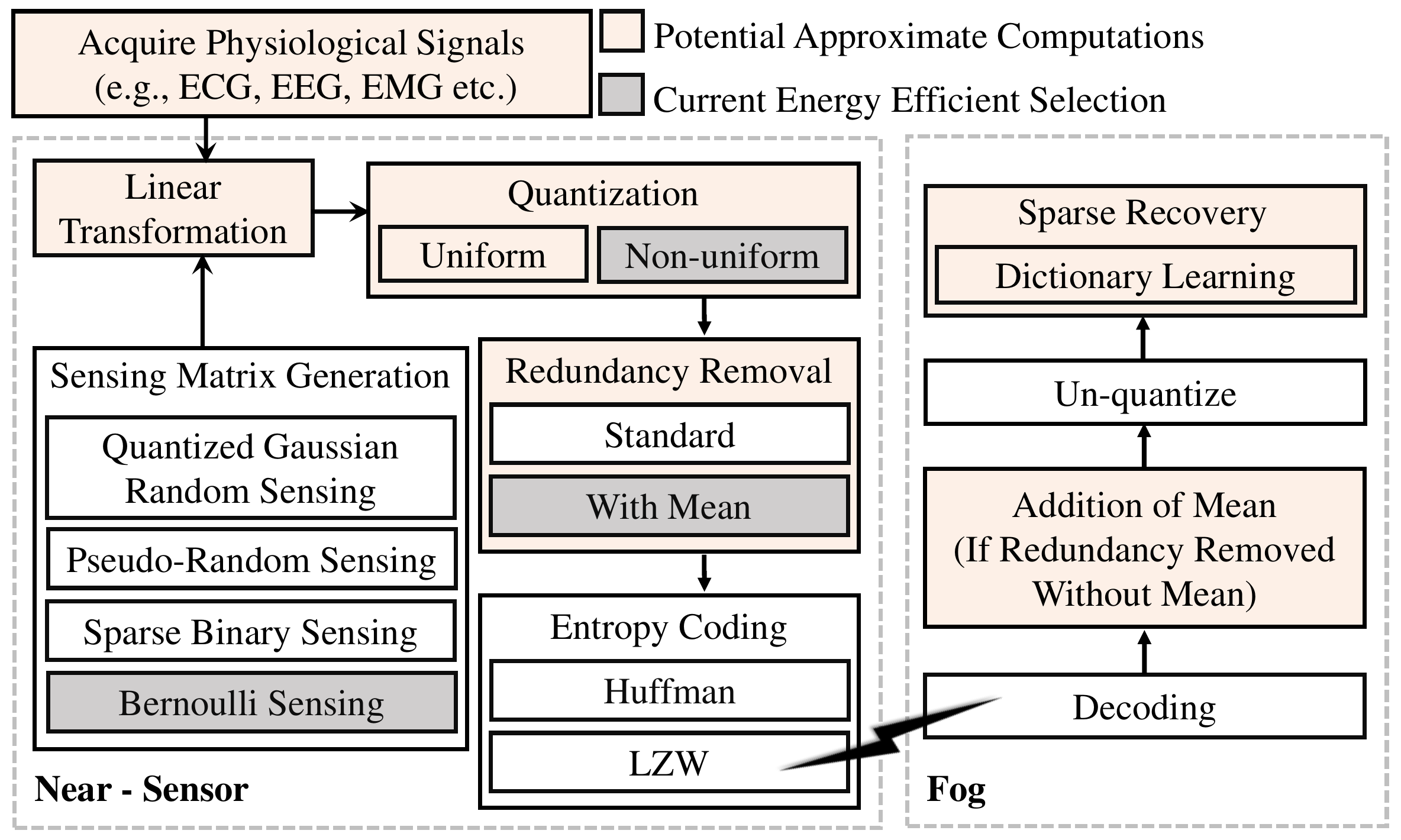}
	\caption{System Architecture of Energy Efficient Digital CS for Hybrid-IoT based WBAN}
	\label{fig:systemoverview}
	\vspace{-0.1in}
\end{figure}

The bio-signals, usually, contain redundant information and can be compressed using Nyquist based lossy compression techniques such as embedded zero tree wavelet transform (EZW)\cite{hilton1997wavelet} and the set partitioning in hierarchical trees (SPIHT) \cite{lu2000wavelet}. These methods involve storage, sorting and complex matrix-vector multiplications which subsequently drain the battery. In the recent past, Compressed Sensing (CS) has emerged as their low power substitute. It exploits the sparsity of the bio-signals and directly acquires the data at a rate proportional to the number of measurements, using sensing matrices. The sensing matrix can be generated through different techniques, like quantized Gaussian random, pseudo-random sensing \cite{mamaghanian2011compressed}, binary antipodal CS (SCS: standard sensing matrix with binary antipodal entries), rankness CS  (RCS), and zeroing CS (ZCS) \cite{mangia2017zeroing}. However, Bernoulli Sensing (BCS) outperforms with significant savings in transmission energy and computational cost \cite{craven2016energy}. Its non-zero Bernoulli distributed elements are usually placed column-wise but our analysis revealed that the row fashion ${\mathrm{BCS}}^\mathrm{*}$ can be opted for less computations. Fig. \ref {fig:comparitiveanalysis2} provides a comparison of (${\mathrm{BCS}}^\mathrm{*}$) with the most commonly used CS acquisition methods in Bluetooth Enabled Environment, using Digital Signal Processing. Since, the main consumer of energy in ambulatory healthcare environments is wireless data transmission so, an extension of CS with quantization, redundancy removal and entropy coding is deemed more advantageous (digital CS) \cite{craven2016energy}. Fig. \ref{fig:systemoverview} presents the system architecture of digtial CS. Although, it results into curtailed wireless transmission cost but the earned on-board energy efficiency is not adequate due to successive compression algorithms.

%%%%%%%%%%%%%%%%%%%%%%%%%%%%%%%%%%%%%%%%%%%%%%%%%%%%%%%%%%%%%%%%%%%%%%%%%%%%%%%%%

\subsection{Research Challenges}
The main research challenges in developing a novel approximate WBAN for IoT-Healthcare Systems are:
\begin{enumerate}
	\item \textbf{Error Margin Exploration:} How to ensure the validity of a computationally energy-aware CS algorithm?
	\item \textbf{Accurate Diagnosis:} How to incorporate the parametric behavior (quality and energy consumption) in approximate CS acquisition (approxCS) to ensure an accurate Arrhythmia detection?
	\item \textbf{Knob Controllability:} Which context aware knobs can be utilized in developing an energy efficient hardware implementation of CS? 
	
\end{enumerate}		

%%%%%%%%%%%%%%%%%%%%%%%%%%%%%%%%%%%%%%%%%%%%%%%%%%%%%%%%%%%%%%%%%%%%%%%%%%%%%%%%%

\subsection{Novel Contributions}
This paper makes the following contributions:
\begin{enumerate}
	\item A \textbf{Generic Methodology} for ApproxCS with design space exploration and thus, enabling its multiple energy-aware architectural versions.
	\item A \textbf{Comprehensive Analysis} of ApproxCS, using context-aware knobs in terms of quality and energy.
	\item \textbf{ASIC Implementation} of CS acquisition by employing accurate and low-power approximate adders (LPAA).
\end{enumerate}

%%%%%%%%%%%%%%%%%%%%%%%%%%%%%%%%%%%%%%%%%%%%%%%%%%%%%%%%%%%%%%%%%%%%%%%%%%%%%%%%%

\subsection{Paper Organization}
The approximate computing in digtal CS is discussed for IoT based WBAN in Section \ref{sec:approxCS}. Section \ref{sec:moti} presents a case study for the motivation towards approximate computing in CS. In Section \ref{sec:methodology}, a generic methodology for approxCS is proposed. Section \ref{sec:CS} discusses the proposed approxCS model along with accurate reconstruction. Section \ref{sec:exp} presents the experimental setup and tool flow of this framework. It also discusses the performance evaluation of approxCS for Arrhythmia diagnosis. Finally, Section \ref{sec:conclusion} concludes the paper.   

\section{Approximate Digital Compressed Sensing}
\label {sec:approxCS}
Approximate computing is a promising approach for energy savings, especially, in conjunction with other state-of-the-art energy efficient techniques \cite {wang2015energy} -\cite{dazzi2018sub}. It exploits the inherent error resilience of healthcare applications and relaxes the equivalence margin between specification and implementation \cite{samie2016iot}.  In correspondence with the `digital CS' model \cite{craven2016energy}, the approximate computing in the basic computational phases of IoT based WBAN is discussed below.

%%%%%%%%%%%%%%%%%%%%%%%%%%%%%%%%%%%%%%%%%%%%%%%%%%%%%%%%%%%%%%%%%%%%%%%%%%%%%%%%%

\subsection{Data Acquisition}
The quality of the input data can be reduced by acquiring the bio-signals at a low sample rate under a low power budget\cite{samie2016iot}. Since most bio-signals are sparse, CS can also be used to reduce the volume of data \cite{bortolotti2014approximate}.

%%%%%%%%%%%%%%%%%%%%%%%%%%%%%%%%%%%%%%%%%%%%%%%%%%%%%%%%%%%%%%%%%%%%%%%%%%%%%%%%%

\subsection{Data Processing}
Through inexact arithmetic units (e.g., adder, multiplier, divider, etc.) and stage skipping, the energy efficiency of certain computations can be improved. For instance, the accurate additions or multiplications in linear transformation, redundancy removal (with mean) and sparse recovery can be replaced with their approximate counterparts. In recursive recovery algorithms, certain loops can also be skipped with a slight accuracy loss \cite{chippa2013analysis}.

%%%%%%%%%%%%%%%%%%%%%%%%%%%%%%%%%%%%%%%%%%%%%%%%%%%%%%%%%%%%%%%%%%%%%%%%%%%%%%%%%

\subsection{Data Storage}
For ultra-low power designs, it is also possible to reduce the storage sizes (memories) or the number of memory accesses. Bortolotti et al. \cite{bortolotti2014approximate} proposed an ultra-low power bio-signal processor through approximate memory units in CS.

%%%%%%%%%%%%%%%%%%%%%%%%%%%%%%%%%%%%%%%%%%%%%%%%%%%%%%%%%%%%%%%%%%%%%%%%%%%%%%%%%

\subsection{Data Transmission}
The approximate computing has been misleadingly used in certain compression algorithms as optimized lossy transmission. However, the approximate arithmetic units can be embedded in the compression algorithms. It will help in saving the computation energy but the transmission energy will remain unaffected due to same number of total computational bits as in an accurate design. 

\begin{figure*}[!ht]
	\centering
	\vspace{-0.1in}
	\includegraphics[width=\textwidth]{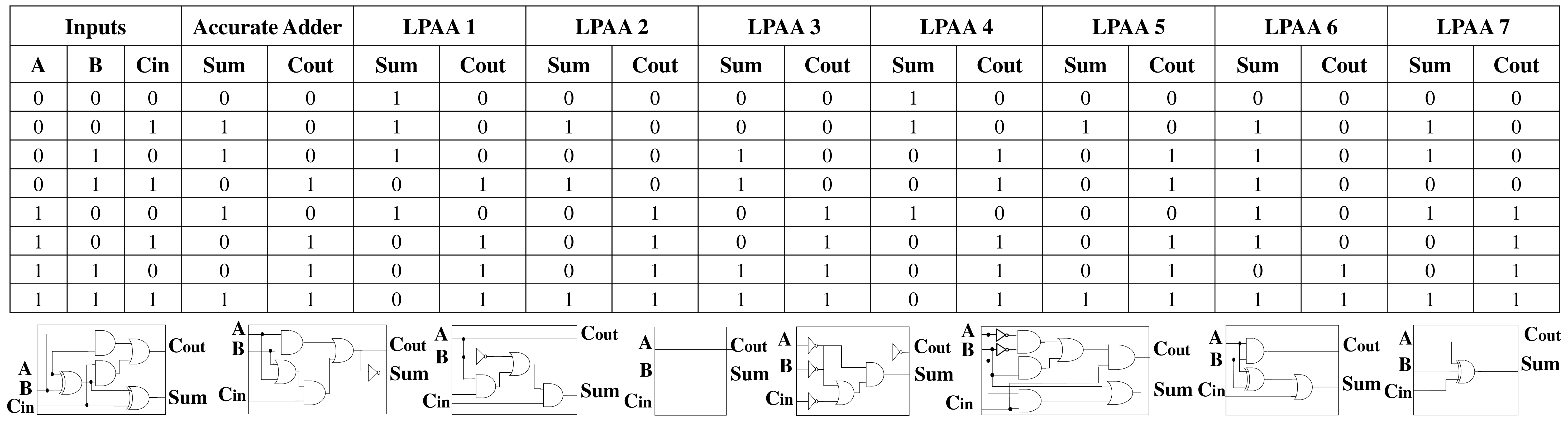}
	\caption{Truth tables and gate-level models of Low-Power Approximate Adders (LPAA) as per the designs in \cite{rehman2016architectural}}
	\label{fig:approx_add}
	\vspace{-0.1in}
\end{figure*}

\begin{figure}[!b]
	\centering
	\vspace{-0.1in}
	\includegraphics[width=1\linewidth]{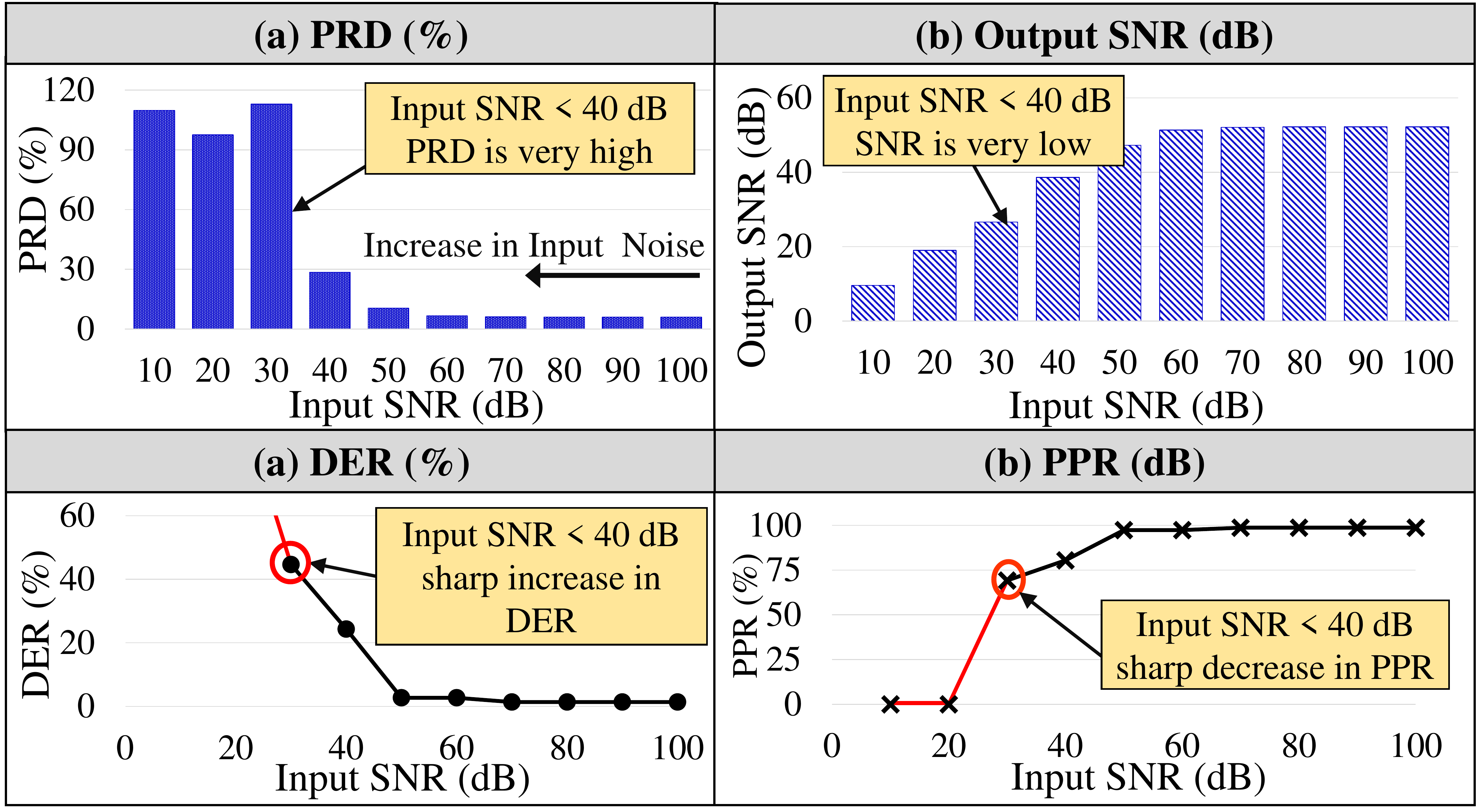}
	\caption{Error Margin Analysis of the ${\mathrm{BCS}}^\mathrm{*}$ using SNR = - 20 $\log ({\frac{||\textbf{x} -\  \mathrm{\textbf{x}}^\mathrm{*}||}{||\textbf{x}||}})$, PRD = ${\frac{||\textbf{x} -\  \mathrm{\textbf{x}}^\mathrm{*}||}{||\textbf{x} -\  \bar{\textbf{x}}||}}$, DER = ${\frac{\mathrm{FP +\ FN}}{\mathrm{TP}}}100 \%$ and PPR = ${\frac{\mathrm{TP}}{\mathrm{TP} +\ \mathrm{FP}}}100\%$ where TP = correctly identified R-peaks, FP = erroneous R-peaks and FN = missed R-peaks. The red line indicates highly distorted and non-recoverable signal}
	\label{fig:ErrorMargin_Comp_8Aug}
	\vspace{-0.1in}
\end{figure}

In this paper, we exploited the near-sensor approximate computing in ${\mathrm{BCS}}^\mathrm{*}$ (approxCS) at the hardware level for ultra-low power acquisition and thereby, explored processing and transmission phases as well. Approximate computing is application specific so, the performance of approxCS is evaluated for Arrhythmia detection, using the most commonly used MIT-BIH Arrhythmia database \cite{Physionet}.

\section{Motivational Case Study: Error Margin Analysis of Compressed Sensing}
\label {sec:moti}
\begin{figure}[!h]
	\centering
	\vspace{-0.1in}
	\includegraphics[width=1\linewidth]{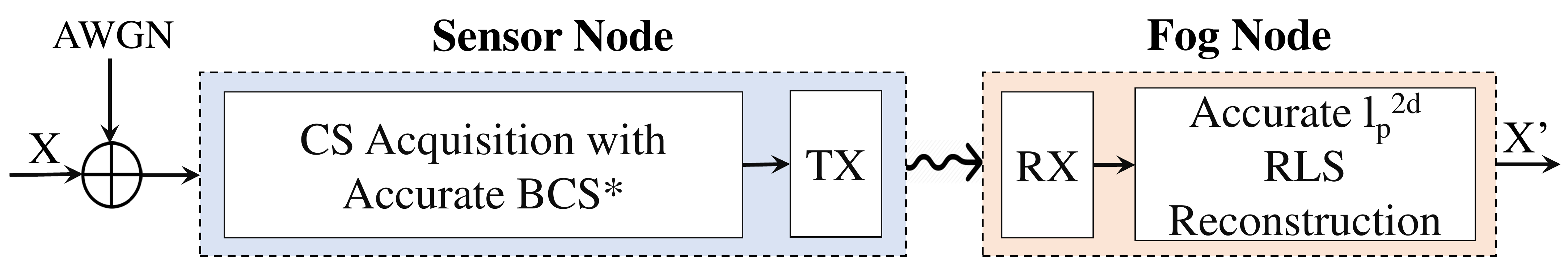}
	\caption{Experimental Setup for Error Margin Analysis of ${\mathrm{BCS}}^\mathrm{*}$}
	\label{fig:Error_exp}
	\vspace{-0.1in}
\end{figure}

To validate the applicability of approximate computing, we first explored the error margin in ${\mathrm{BCS}}^\mathrm{*}$ with additive white Gaussian noise (AWGN) and an outperforming ${\mathrm{l}_\mathrm{p}}^\mathrm{2d}$ RLS algorithm \cite{pant2014compressive}. As shown in Fig. \ref{fig:Error_exp}, the noise is added to the input ECG signal. The performance is evaluated for Arrhythmia diagnosis (patient record: 100m) using four well-adopted performance metrics: Signal to Noise Ratio (SNR) and Percentage Root Difference (PRD) for measuring the distortion in the signal, and Positive Predictive Rate (PPR) and Detection Error Rate (DER) for quantifying Arrhythmia detection. Fig. \ref{fig:ErrorMargin_Comp_8Aug} shows that an increase in AWGN noise causes high PRD and DER and low SNR and PRD (undesirable) and thus, high chance of missing the critical events. Moreover, there exists significant error tolerance (20\%-30\%) and thus, huge potential for approximations.

In this paper, we used seven advanced state-of-the-art 1-bit LPAA \cite{rehman2016architectural} with ripple carry chain for multiple bits. Figure \ref{fig:approx_add} refers to their gate-level designs and truth tables. The existing low-latency adders \cite{shafique2015low}\cite{hanif2017quad} experience carry chain breaking and may lead to undesirable errors. Our approxCS methodology is orthogonal to the use of any type of approximate adder.

%%%%%%%%%%%%%%%%%%%%%%%%%%%%%%%%%%%%%%%%%%%%%%%%%%%%%%%%%%%%%%%%%%%%%%%%%%%%%%%%%%%%%%%%%%%%%%%%%

\section{Proposed Methodology}
\label{sec:methodology}
\begin{figure}[!h]
	\centering
	\vspace{-0.1in}
	\includegraphics[width=1.0\linewidth]{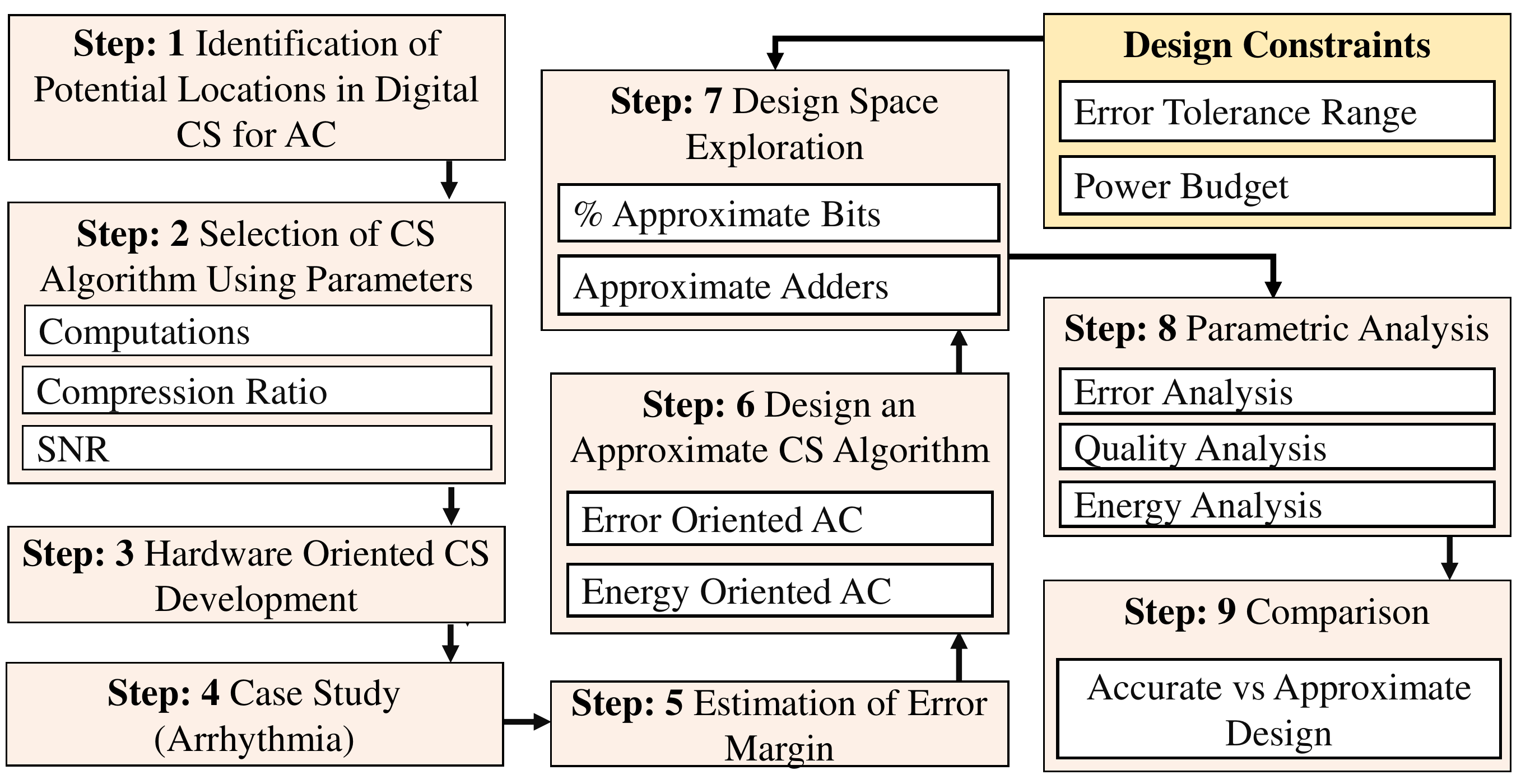}
	\caption{Proposed Methodology for Applying Approximate Computing in Compressed Sensing Algorithms}
	\label{fig:methodology}
	\vspace{-0.1in}
\end{figure}

An overview of the proposed methodology for designing ApproxCS is presented in Fig. \ref {fig:methodology}. It consists of the following four key steps:

\begin{enumerate}
	\item \textbf{Error Margin Analysis:} Since, approximate computing is itself an error inducing practice so, the first step is to identify the error margin in the healthcare application. This can be achieved by assuming an input signal corrupted with AWGN and then, analyzing its impact in comparison to a noise-free algorithm (see Fig. \ref{fig:Error_exp}). The addition of noise in the compressed stream can be crucial and can lead to incorrect identification of R-peaks. However, if a comparatively low degradation in the output quality (e.g., SNR) is found and the number of mis-predicted critical events are found to be below a certain threshold as prescribed by the diagnostics expert, then the approximate computing is applicable. 
	
	\item \textbf{Hardware Error Resilience Analysis:} There are two possible hardware approximation knobs that can be exploited for improving the energy efficiency. First, the ‘Fixed-point Quantization’ is carried out by transforming the floating-point operations to fixed point and reducing the word sizes. Later, ‘approximate hardware units’ are used for approxCS modeling. 
	
	\item \textbf{Design Space Exploration:} An exploration heuristic is employed that selects a subset of energy-efficient design points like approximate number of bits and type of LPAA while considering the quality constraint and the available energy budget. 
	
	\item \textbf{Training for Performance Improvement:} The reconstruction algorithm is then trained with a dictionary learning (DL) algorithm \cite{pant2014compressive} for better reconstruction quality and the design points are re-evaluated in the approxCS model.
\end{enumerate}

\section{Approximate Computing based Compressed Sensing Framework}
\label{sec:CS}
The basic CS model includes data compression and reconstruction. The signal is first  compressed in the data acquisition phase before transmitting it over a Gaussian (or any) channel. The channel induces AWGN in the system  and its effect can be compensated with a reconstruction algorithm at the receiver end. In the following section, an energy efficient approxCS algorithm, while considering a noisy environment, has been developed and it mainly targets the ECG signals. The approxCS only contributes towards computation energy savings, but the wireless transmission energy and cost does not vary. This is due to the fact that the size of the received signal and the number of transmission bits remains the same as in an accurate model.   

%%%%%%%%%%%%%%%%%%%%%%%%%%%%%%%%%%%%%%%%%%%%%%%%%%%%%%%%%%%%%%%%%%%%%%%%%%%%%%%%%

\subsection{Signal Acquisition}
Generally, in CS signal acquisition, the sparse input signal \textbf{x} of length \textit{N} is acquired through its real-time linear transformation with a sensing matrix $\mathbf{\Phi}$ of size \textit{NxM} (such that \textit{M}$<<$\textit{N}) to obtain a compressed measurement vector \textbf{y} of length \textit{M}. Mathematically, this relation can be written as: 
\begin{equation}\mathbf{y}\ =\mathbf{\Phi x}+\mathbf{w}\end{equation}   

\noindent where \textbf{w} represents the AWGN channel noise. To circumvent the on-board generation of the sensing matrices, the concept of \textit{sparse multiplier} is exploited in linear transformation. A sorted vector \textbf{z}, specifying the locations of 'r' ones in each row of the sensing matrix, is used for successive approximate additions of the corresponding ECG signal values. As this process does not summon a sorting algorithm and matrix multiplications so, it is assumed to be less time consuming and thus, more feasible for embedded systems. The approximate additions, for each element in \textbf{y}, can be written as: 

\begin{equation}\mathbf{y}_\mathrm{k,\textit{p}}\ = \mathbf{y}_\mathrm{k-1, \textit{p}} +\ \sum_{\mathrm{\textit{q}}} \mathbf{x} (\mathbf{z}_\mathrm{\textit{q}}) \end{equation}		  

\noindent where 0 $\le$ \textit{q} $<$ \textit{r} for p = 0 and r $\le$ \textit{q} $<$ \textit{2r} for p $\neq$ 0.

%%%%%%%%%%%%%%%%%%%%%%%%%%%%%%%%%%%%%%%%%%%%%%%%%%%%%%%%%%%%%%%%%%%%%%%%%%%%%%%%%

\subsection{Signal Recovery}
In the bio-signals, there exists a temporal correlation between samples which can be exploited for reducing the approximation noise. Due to this correlation, their first order difference is usually sparser than the signal itself. Consequently, the method which improves sparsity on the first order difference can yield improved performance as compared to the algorithms which promote sparsity on the signal itself. In \cite{pant2014compressive}, Pant et al. proposed a sequential version `${\mathrm{l}_\mathrm{p}}^\mathrm{2d}$ Recursive Least Squares (RLS) algorithm' of the basic conjugate-gradient algorithm, which yields an improved reconstruction performance (higher SNR at low compression rates), as compared to other state-of-the-art algorithms \cite{pant2014compressive2} \cite{pant2011unconstrained}. The ${\mathrm{l}_\mathrm{p}}^\mathrm{2d}$ RLS algorithm recovers the sparse signal $\mathrm{\textbf{x}}^\mathrm{*}$ by minimizing the lp second order difference pseudo-norm, ${||2\mathrm{d\mathbf{\textbf{x}}}\||}_\mathrm{p}^\mathrm{p}$, and solving the following p-RLS problem.

\begin{equation}\min_{\textbf{x}}{\ \ \ \ \ \ \ f\left(x\right)=\ \frac{1}{2}} {||\mathbf{\Phi} \textbf{x} -\ \textbf{y}\||}_2^2 + \lambda\ {||2\mathrm{\textbf{d}\textbf{x}}\||}_\mathrm{p}^\mathrm{p,\epsilon}\end{equation} 

where $\lambda$ is the regularization parameter and $\epsilon$ is the approximation parameter which renders the function ${||2\mathrm{d\textbf{x}}\||}_\mathrm{p}^\mathrm{p}$ smooth. In this paper, the signal recovery has accurate implementation for the evaluation of approxCS. 

\section{Results and Discussions}
\label{sec:exp}

\subsection{Experimental Tool Flow}
\begin{figure}[!h]
	\centering
	\vspace{-0.1in}
	\includegraphics[width=1\linewidth]{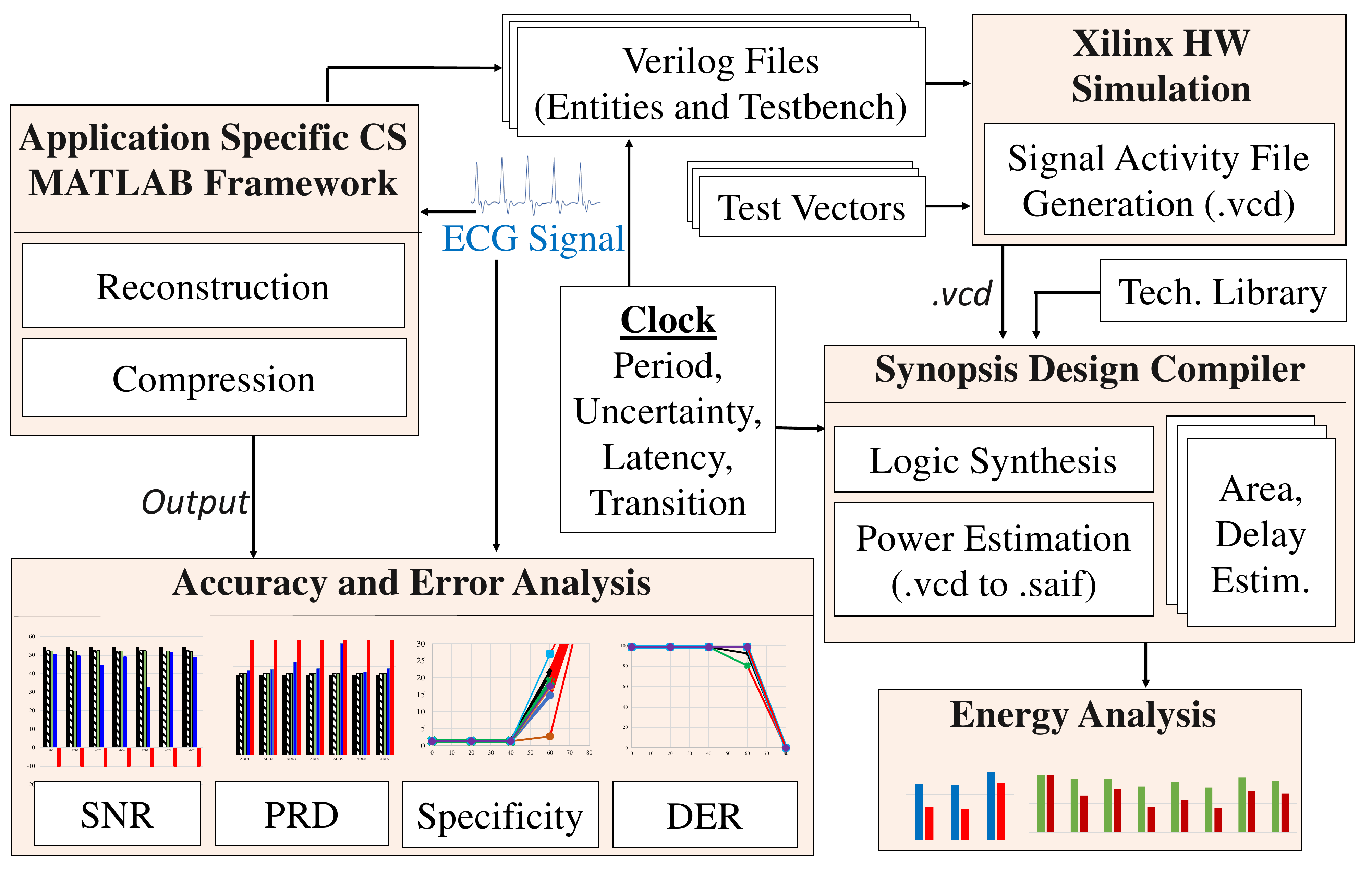}
	\caption{Experimental Tool Flow}
	\label{fig:ExperimentalToolFlow}
	\vspace{-0.1in}
\end{figure}

For simulations and design tools, we use an Intel Core i7-6700T Quad-Core server operating at 3.06 GHz with 32 GB of RAM. Fig. \ref{fig:ExperimentalToolFlow} presents our integrated tool flow. According to this setup, the accuracy of approxCS framework is first evaluated for Arrhythmia detection in MATLAB and then, its corresponding behavioral Verilog code is developed. After confirming the correct behavior of the hardware in Verilog, Value Change Dump (VCD) file is generated and fed into the Synopsys Primetime tool to generate latency and power reports using a 65nm technology library. 

%%%%%%%%%%%%%%%%%%%%%%%%%%%%%%%%%%%%%%%%%%%%%%%%%%%%%%%%%%%%%%%%%%%%%%%%%%%%%%%%%

\subsection{System Design Configurations}
\begin{figure}[!t]
	\centering
	\vspace{-0.1in}
	\includegraphics[width=1\linewidth]{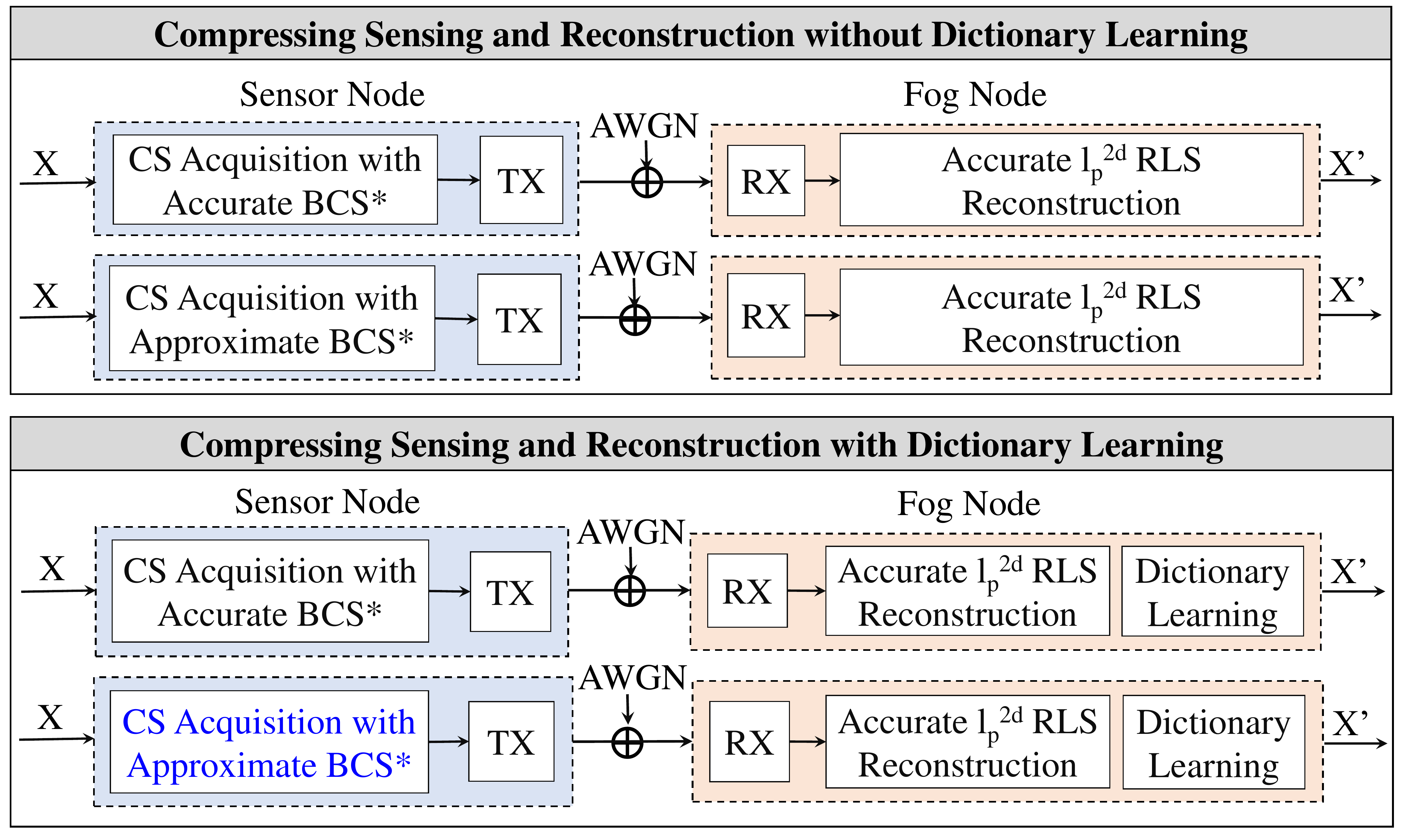}
	\caption{Experimental Setup for the comparison between Accurate and Approximate CS with dictionary independent and dependent accurate reconstruction}
	\label{fig:experiments}
	\vspace{-0.1in}
\end{figure}

The basic CS model is usually very compute-intensive due to alarge vectors or matrices. It takes on average about 22-24 hours for 10 MATLAB simulations in case of Arrhythmia detection (1 minute recording of patient record 100m). To expedite the execution of this algorithm, the least possible configurations for the sensing matrix, measurement vector and other parameters are first explored through extensive testing of the accurate CS acquisition with reliable reconstruction. With this viewpoint, the \textit{sensing matrix} of size 256x128, is constructed such that any two entries in its each row are randomly set to unity. Such sensing matrices, with different configurations, have recently been used in \cite{pant2011unconstrained}, \cite{pant2014compressive}, and \cite{craven2016energy}. The measurement of compressed signal is obtained by linearly transforming the input ECG signal, having 256 measurements, with this measurement matrix (using sparse multiplier concept). Furthermore, the parameters, in the \textit{recovery algorithm} \cite{pant2014compressive}, are initialized as: $\mathrm{\epsilon}_\mathrm{1}$ = 1, $\mathrm{\lambda}_\mathrm{1}$ = 1, T = 50, $\mathrm{E}_\mathrm{t}^\mathrm{-15}$, $\mathrm{L}_\mathrm{b}$ = 15, r = 4 and $\delta$ = $\mathrm{10}_\mathrm{-5}$. The \textit{DL} algorithm \cite{pant2014compressive} is used with the same parameters except T = 5. Fig. \ref{fig:experiments} presents our four experimental configurations with AWGN channel noise of zero mean and variance $\mathrm{4x10}^\mathrm{-4}$ \cite{pant2014compressive}.

%%%%%%%%%%%%%%%%%%%%%%%%%%%%%%%%%%%%%%%%%%%%%%%%%%%%%%%%%%%%%%%%%%%%%%%%%%%%%%%%%

\subsection{Selection of Dataset for Arrhythmia Detection}

For the validation of the proposed methodology for Arrythmia diagnosis, the approxCS is realized with the MIT-BIH Arrhythmia database \cite{Physionet}. This dataset contains 48 patient records with modified limb lead II (MLII) as the principal lead. Each recording is sampled at 360 Hz with a 11 bit resolution. 

In this paper, the performance of approxCS is quantified for substantiating R-peak detection in `Tachycardia and Bradycardia' using four performance metrics (mentioned in Fig. \ref{fig:ErrorMargin_Comp_8Aug}). 

\begin{figure}[!h]
	\centering
	\vspace{-0.1in}
	\includegraphics[width=1\linewidth]{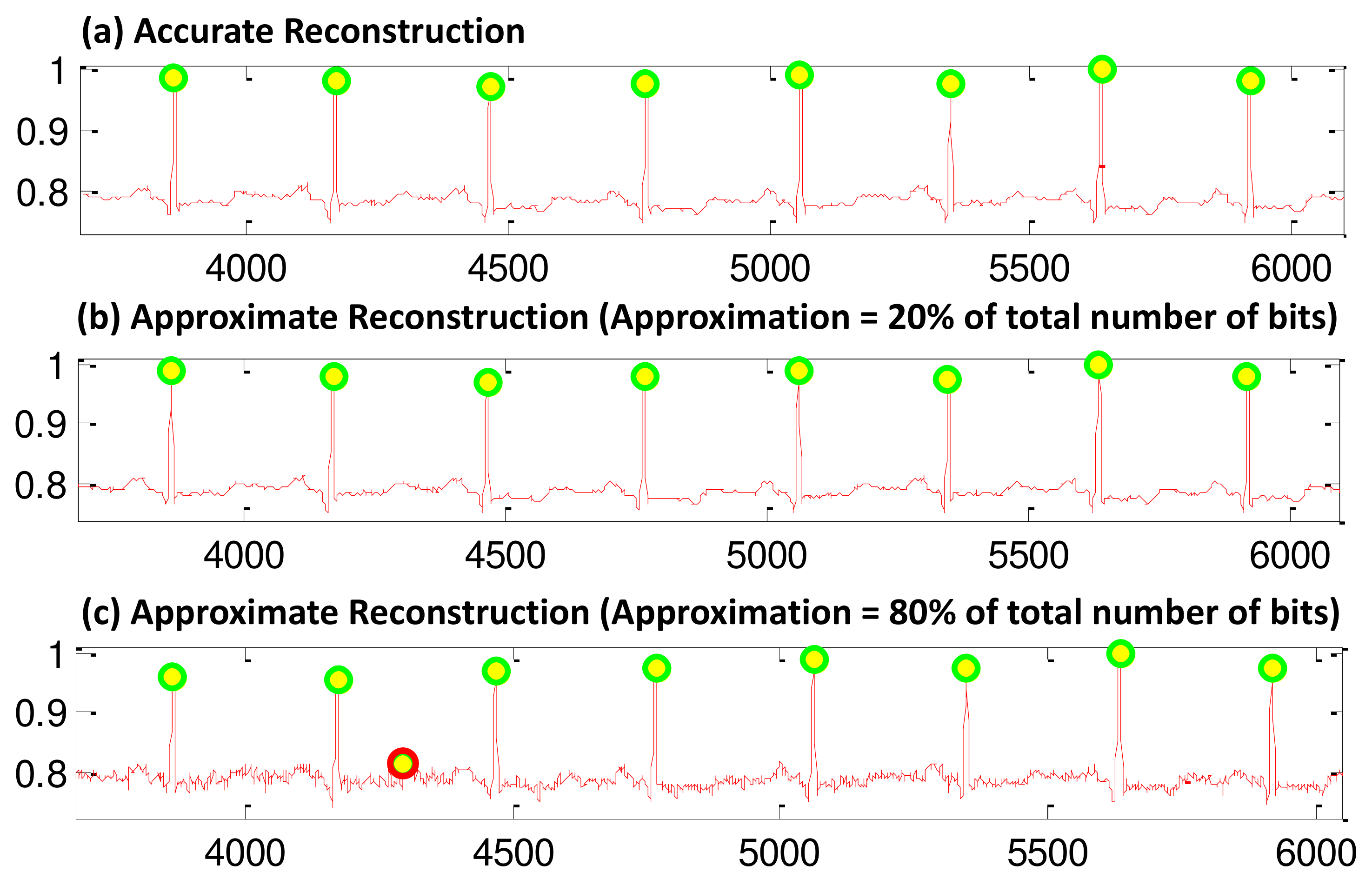}
	\caption{Effect of Approximation Noise on Arrhythmia Detection (patient record: 100m): The green and red circles indicate TP and FP R-peaks respectively.}
	\label{fig:ecg_signals}
	\vspace{-0.2in}
\end{figure}

%%%%%%%%%%%%%%%%%%%%%%%%%%%%%%%%%%%%%%%%%%%%%%%%%%%%%%%%%%%%%%%%%%%%%%%%%%%%%%%%%

\subsection{Performance Evaluation}
\label{subsec:perf}

The accurate ${\mathrm{l}_\mathrm{p}}^\mathrm{2d}$ RLS reconstruction algorithm \cite{pant2014compressive} uses normalized ECG signals and hence, the corresponding CS acquisition is based on floating point computations. However, FPGAs and ASICs support fixed point computations only. As, the fixed point quantization results into truncation of some bits so, CS compression algorithm is first analyzed by varying the fractional bits. However, the decimal bits are set as per the requirement of each computation and kept constant. It is observed that the reconstruction performance with 33 and 43 fractional bits in  CS compression and reconstruction, respectively, seems to be the best optimal solution with transmission energy savings and without quality degradation. In this paper, different percentages of total bits have been used for employing approximate computing. 

\begin{figure}[!b]
	\centering
	\vspace{-0.1in}
	\includegraphics[width=1\linewidth]{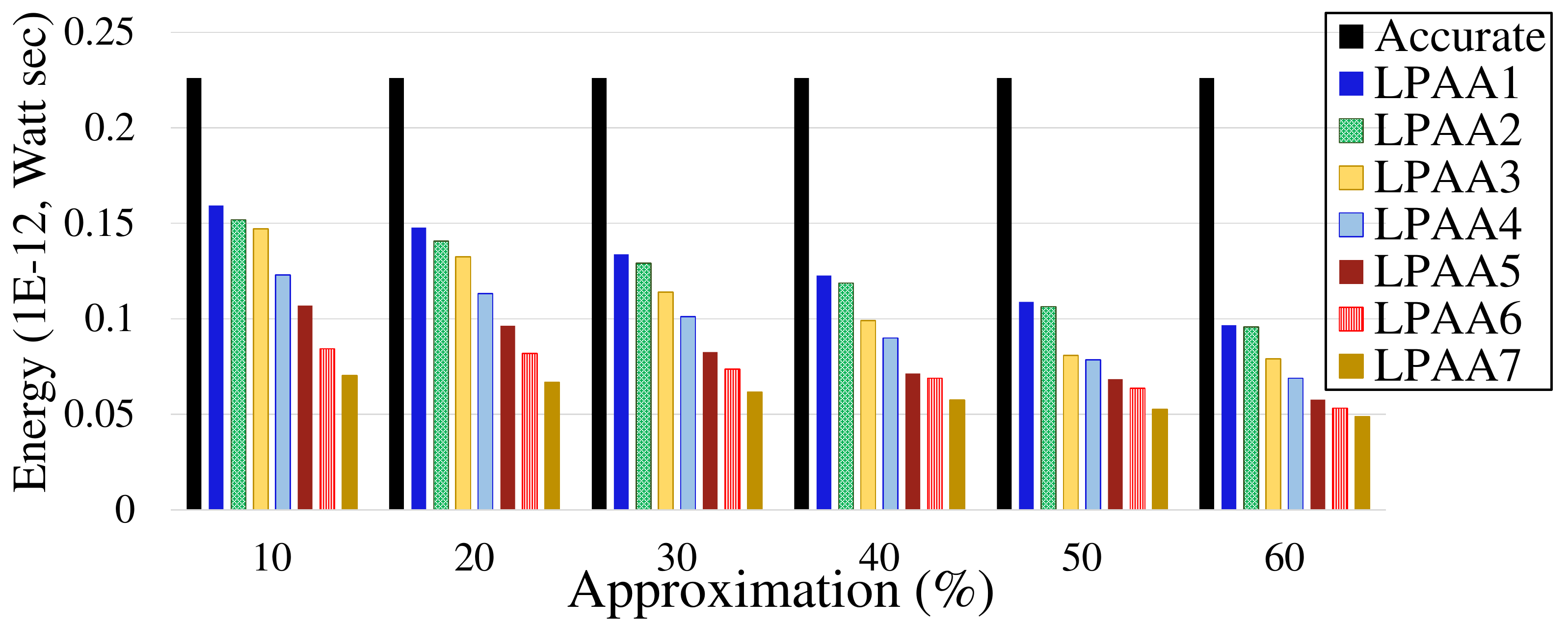}
	\caption{Energy Consumption of Accurate and Approximate ${\mathrm{BCS}}^\mathrm{*}$}
	\label{fig:Compression_Energy}
	\vspace{-0.2in}
\end{figure}

\begin{figure*}[!t]
	\centering
	\vspace{-0.1in}
	\includegraphics[width=1\linewidth]{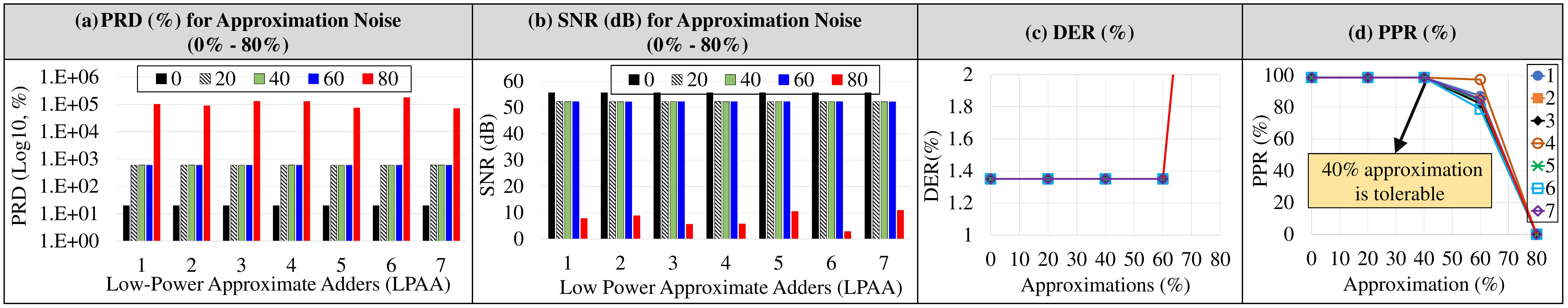}
	\caption{Performance Evaluation of 'Approximate ${\mathrm{BCS}}^\mathrm{*}$ with ${\mathrm{l}_\mathrm{p}}^\mathrm{2d}$ RLS Reconstruction' for Arrhythmia Detection (1 min recording of patient record 100m). Red bars show unrecoverable signals}
	\label{fig:Results_reconstruction1}
	\vspace{-0.1in}
\end{figure*}

\begin{figure*}[!h]
	\centering
	\vspace{-0.01in}
	\includegraphics[width=1\linewidth]{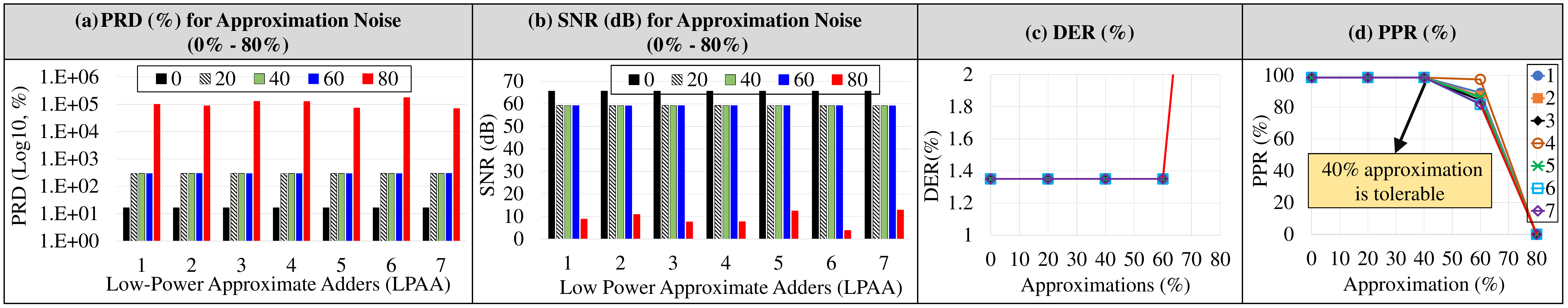}
	\caption{Performance Evaluation of 'Approximate ${\mathrm{BCS}}^\mathrm{*}$ with DL based ${\mathrm{l}_\mathrm{p}}^\mathrm{2d}$ RLS Reconstruction' for Arrhythmia Detection (10\% and 90\% training and test data from full length recording of patient record 100m). Red bars show unrecoverable signals)}
	\label{fig:Results_reconstruction2}
	\vspace{-0.2in}
\end{figure*}

%%%%%%%%%%%%%%%%%%%%%%%%%%%%%%%%%%%%%%%%%%%%%%%%%%%%%%%%%%%%%%%%%%%%%%%%%%%%%%%%%

\subsection{Design Space Exploration and Parametric Analysis}

Approximating the CS results into noisy R-peaks. However, the approximation noise grows from the baseline of the ECG signal (see Fig. \ref{fig:ecg_signals}) so, the correct R-peaks are never missed (FN = 0). The effect of approximate computing on CS acquisition, alongwith accurate ${\mathrm{l}_\mathrm{p}}^\mathrm{2d}$ RLS reconstruction dependent and independent of DL, is illustrated in Figs. \ref{fig:Results_reconstruction1} and \ref{fig:Results_reconstruction2}. The summary of these results is provided below:

\begin{enumerate}
	\item The 80\% approximation noise is too much for the reconstruction algorithm (very high PRD and low SNR) and thus, the ECG signal is not recoverable.    
	\item Upto 40 \% approximation, all LPAA exhibits low degradation in signal quality (SNR) but LPAA 7 has the lowest energy consumption (see Fig. \ref{fig:Compression_Energy}).
	
	\item LPAA 4 exhibits comparatively better reconstruction with even 60\% approximation but some R-peaks are noisy. So, this knob can only be considered at the cost of slight accuracy loss in Arrhythmia detection.   
	
	\item With 60 \% approximation noise, LPAA 6 has very low PPR, i.e., very high probability of noisy R-peaks.
	
	\item According to Fig. \ref{fig:Compression_Energy}, an increase in the number of approximate bits causes more energy savings. With tolerable approximation error, LPAA 7 has the minimum energy consumption (see Fig. \ref{fig:Compression_Energy}). It saves approximately 59\% energy with very low probability of incorrectly detected R-peaks.  
	
	\item 1-40\% approximation does not strongly affect the signal quality and such approxCS design can be opted for energy efficiency in ${\mathrm{BCS}}^\mathrm{*}$ and Arrhythmia diagnosis.  
	
	\item In general, the DL algorithm enhances the reconstruction quality. In Fig \ref {fig:Results_reconstruction2}, an improved SNR and PPR and low DER and PRD can be clearly observed with LPAA 7.
	
	\item The DL algorithm is used on the receiving end so, the computation energy savings remain the same at the near-sensing node.    
\end{enumerate}	

\section{Conclusion}
\label{sec:conclusion}
In this paper, we exploited the inherent sensing noise resilience of the bio-signals by applying the approximate computing in an accurate ${\mathrm{BCS}}^\mathrm{*}$ acquisition for improving the energy efficiency of near-sensing node. In particular, we identified multiple application-aware approximation knobs using state-of-the-art seven low-power approximate adders. For illustration, we demonstrated the trade-off between energy efficiency and the output quality of Arrhythmia detection and thereby, obtained approximately 59\% energy savings. \textit{In real-world applications, it is not necessary to utilize all computational resources of near-sensing node for Compressed Sensing but the energy consumption can be reduced by employing approximate computing and designing an energy efficient healthcare device depending on the error tolerance of the considered healthcare application.}

%========================================================================================================
%\small
%\footnotesize
\bibliographystyle{IEEEtran}
\bibliography{bib/conf}

\end{document}